\documentstyle[amssymb,amsmath,12pt]{elsart}

\def\ee{\end{equation}}
\def\be#1{\begin{equation}\label{#1}}

\def\ba{\begin{array}}
\def\ea{\end{array}}

\begin{document}

\begin{frontmatter}
\title{{Consistent deformations method applied to a topological coupling
of antisymmetric gauge fields in D=3}}
\author[UFC,UECE]{D. M. Medeiros},
\author[UFC]{R. R. Landim} and
\author[UFC]{C. A. S. Almeida\thanksref{e-mail}}
\address[UFC]{ Universidade Federal do Cear\'{a} - Departamento de
F\'{\i}sica \\ C.P. 6030, 60470-455 Fortaleza-Ce, Brazil}
\address[UECE]{ Universidade Estadual do Cear\'{a} - Departamento de
F\'{\i}sica e Qu\'{\i}mica \\Av. Paranjana, 1700, 60740-000
Fortaleza-Ce, Brazil}
\thanks[e-mail]{Electronic address: carlos@fisica.ufc.br}

\begin{abstract}
In this work we use the method of consistent deformations of the
master equation by Barnich and Henneaux in order to prove that an
abelian topological coupling between a zero and a two form fields
in D=3 has no nonabelian generalization. We conclude that a
topologically massive model involving the Kalb-Ramond two-form
field does not admit a nonabelian generalization. The introduction
of a connection-type one form field keeps the previous result.

\end{abstract}
\end{frontmatter}



\vspace{1.0cm}

Antisymmetric tensor gauge fields coupled to abelian or nonabelian
gauge fields has been studied during the last decade. In
particular, four-dimensional BF models has been extensively
investigated in different contexts (see Ref. \cite{blau} and
references therein). Besides supergravity and string theories
\cite{schwarz}, two form fields play an important role in
topological mass generation mechanism to Kalb-Ramond fields
\cite{la1,gener}.  Indeed, a nonabelian theory involving an
antisymmetric tensor field coupled to a gauge field appears as an
alternative mechanism for generating vector bosons masses, similar
to the theory of a heavy Higgs particle \cite{freedman}. Therefore
analysis of nonabelian topological terms deserves some attention.
It is worth to mention a generalization to a compact nonabelian
gauge group of an abelian mechanism in the context of nonabelian
quantum hair on black holes \cite{lahiritwo}.

Using the technique of consistent deformation, Henneaux {\it et
al.}~\cite{no-go}, have proved that is not possible to generalize
the four dimensional topological mass mechanism to its nonabelian
counterpart with the same field contents and fulfilling the
power-counting renormalization requirements. In this way, they put
in more rigorous grounds the need to add an auxiliary field.

Let us begin by presenting a recently proposed abelian
three-dimensional action with a topological term involving a
two-form gauge field $B_{\mu \nu }$ and a scalar field $\varphi $
\cite{non-chern}:
\begin{equation}
S_{inv}^A=\int \!\!d^3x~\left( \frac 1{12}H_{\mu \nu \alpha }H^{\mu \nu
\alpha }+\frac 12\partial _\mu \varphi \partial ^\mu \varphi +\frac
m2\epsilon ^{\mu \nu \alpha }B_{\mu \nu }\partial _\alpha \varphi \right) ,
\label{sa}
\end{equation}
where $H_{\mu \nu \alpha }$ is the totally antisymmetric tensor
$H_{\mu \nu \alpha }=\partial _\mu B_{\nu \alpha }+\partial
_\alpha B_{\mu \nu }+\partial _\nu B_{\alpha \mu }$. The action
(\ref{sa}), is invariant under the transformation
\begin{equation}
\delta \varphi =0,\quad \quad \delta B_{\mu \nu }=\partial _{[\mu
}\omega _{\nu ]}~.  \label{invsa}
\end{equation}
The model described by action (\ref{sa}) can be consistently obtained by
dimensional reduction of a four-dimensional $B\wedge F$ model if we discard
the Chern-Simons-like terms \cite{non-chern}.

The purpose of the present note is to prove that there is no power
counting renormalizable nonabelian generalization of the action
(\ref{sa}). We begin making an analysis of the possibility to
construct the nonabelian action with $\varphi \rightarrow \varphi ^a$ and $%
B_{\mu \nu }\rightarrow B_{\mu \nu }^a$, i.e, with the same field
content, and the same number of local symmetries, by making use of
the method of consistent deformation~\cite{def}. As will be
proved, there is a no-go theorem for this construction. The same
occur with an introduction of a connection-type one form gauge
field.

Let us now apply the consistent deformation method described in ~\cite{def}.
We shall start therefore with the following invariant action
\begin{equation}
S_0^{\prime }=\int \!\!d^3x~\left( \frac 1{12}H_{\mu \nu \alpha }^aH^{a\mu
\nu \alpha }+\frac 12\partial _\mu \varphi ^a\partial ^\mu \varphi ^a\right)
,  \label{s0l}
\end{equation}
where now $\varphi ^a$ and $H_{\mu \nu \alpha }^a$ are scalar
fields and the abelian curvature tensor for a set of $N$ fields.
All fields are valued in the Lie algebra ${\cal G}$ of some Lie
group G. Since we are interested if the mass term can exist in a
nonabelian extension of ~(\ref{sa}), the mass parameter will be
considered as a deformation parameter. The action ~(\ref{s0l}) is
invariant under the transformations
\begin{equation}
\delta \varphi ^a=0~,\quad \quad \delta B_{\mu \nu }^a=\partial _\mu \omega
_\nu ^a-\partial _\nu \omega _\mu ^a.  \label{invs0l}
\end{equation}
Since the transformation of $B_{\mu \nu }^a$ is reducible, we introduce a
set of ghosts $(\eta _\mu ^a,\rho ^a)$, where $\eta _\mu ^a$ is a ghost for
the gauge transformation of $B_{\mu \nu }^a$, and $\rho ^a$ the ghost for
ghost for taking into account this reducibility. For all fields of the model
we introduce the corresponding antifields $(B_{\mu \nu }^{*a},\varphi
^{*a},\eta _\mu ^{*a},\rho ^{*a})$. The antifields action reads
\begin{equation}
S_{ant}^{\prime }=\int \!\!d^3x~\left( \frac 12B^{*\mu \nu a}\partial _{[\mu
}\eta _{\nu ]}^a+\eta ^{*\mu a}\partial _\mu \rho ^a\right) .  \label{s0la}
\end{equation}
The free action
\begin{equation}
S_0=S_0^{\prime }+S_{ant}^{\prime },  \label{s0}
\end{equation}
is solution of the master equation
\begin{equation}
(S_0,S_0)=0,  \label{me}
\end{equation}
with
\begin{equation}
(S_0,S_0)=\int \!\!d^3x~\left( \frac{\delta S_0}{\delta \varphi ^a}\frac{%
\delta S_0}{\delta \varphi ^{*a}}+\frac 12\frac{\delta S_0}{\delta B^{a\mu
\nu }}\frac{\delta S_0}{\delta B_{\mu \nu }^{*a}}+\frac{\delta S_0}{\delta
\eta ^{a\mu }}\frac{\delta S_0}{\delta \eta _\mu ^{*a}}+\frac{\delta S_0}{%
\delta \rho ^a}\frac{\delta S_0}{\delta \rho ^{*a}}\right)  \label{eme}
\end{equation}
The nilpotent BRST transformation $s$ on all fields and antifields
is
\begin{equation}
\begin{array}{ll}
s\varphi ^a=0\,, & s\varphi _a^{*}=-\partial ^2\varphi \,, \\
&  \\[1mm]
sB_{\mu \nu }^a=\partial _\mu \eta _\nu ^a-\partial _\nu \eta _\mu ^a\,, &
sB_a^{*\mu \nu }=-\partial _\rho H_a^{\rho \mu \nu }\,, \\
&  \\[1mm]
s\eta _\mu ^a=\partial _\mu \rho ^a\,, & s\eta _a^{*\mu }=\partial _\rho
B_a^{*\rho \mu }\,, \\
&  \\[1mm]
s\rho ^a=0\,, & s\rho _a^{*}=-\partial _\mu \eta _a^{*\mu }\,. \\
&  \\[1mm]
&
\end{array}
\label{brst}
\end{equation}
We show in the table (\ref{gh1-number-dim}) below, the canonical
dimension and the ghost number for all fields and antifields of
the model.
\begin{table}
\begin{center}
\begin{tabular}{|c|c|c|c|c|c|c|c|c|}
\hline
& $\Phi^{a}$ & $B^{a}_{\mu\nu}$ & $\eta^{a}_{\mu}$ & $\rho^{a}$ & $%
\Phi^{\ast a}$ & $B^{\ast a}_{\mu\nu}$ & $\eta^{\ast a}_{\mu}$ &
$\rho^{\ast a}$   \\ \hline $N_{g}$ & 0 & 0 & 1 & 2 & -1 & -1 & -2
& -3  \\ \hline $dim$ & 1/2 & 1/2 & -1/2 & -3/2 & 5/2 & 5/2 & 7/2
& 9/2  \\ \hline
\end{tabular}
\end{center}
\caption{Ghost numbers and dimensions.} \label{gh1-number-dim}
\end{table}

Having the ghost number and dimension of all fields and antifields at hand,
we are now able to solve our problem using the consistent deformation
method. The action ~(\ref{s0}) will be deformed to a new action $S$ in
powers of the deformation parameters:
\begin{equation}  \label{def}
S=S_0+\sum_i g_i S_i+\sum_{i,j}g_i g_j S_{ij}+ \cdots~,
\end{equation}
where $S_i, S_{ij}..$ are local integrated polynomials with ghost number
zero and dimension bounded by three, and $g_i$ are the deformed parameters
with nonnegative mass dimension. The action (\ref{def}) must satisfy the
master equation
\begin{equation}  \label{cons}
(S,S)=0.
\end{equation}
Expanding the master equation (\ref{cons}) in powers of the deformation
parameters, we have
\begin{equation}  \label{order0}
(S_0,S_0)=0,
\end{equation}
\begin{equation}  \label{order1}
(S_0,S_i)=0,
\end{equation}
\begin{equation}  \label{order2}
2(S_0,S_{ij})+(S_i,S_j)=0.
\end{equation}
The equation (\ref{order0}) is the master equation for the $S_0$,
and it does not give any additional information. The equation
(\ref{order1}) tell us that $S_i$ has to be a BRST invariant under
(\ref{brst}). We must neglect BRST exacts, since this correspond
to fields redefinitions. The last equation (\ref{order2}) is
satisfied only if the antibracket $(S_i,S_j)$ is a trivial
cocycle.

Let us now construct all $S_i$ solution of equation (\ref{order1}). First we
focus our attention to terms that do not deform the gauge symmetry, i.e,
terms constructed with the fields only. Due to trivial BRST transformation
of $\varphi ^a$, the all possible terms with this field are
\begin{equation}
S_1=\int \!\!d^3x~\left( \alpha _a\varphi _a\right) ,\quad S_2=\int
\!\!d^3x~\left( \alpha _{ab}\varphi _a\varphi _b\right) ,  \label{phionly12}
\end{equation}
\begin{equation}
S_3=\int \!\!d^3x~\left( \alpha _{abc}\varphi _a\varphi _b\varphi _c\right)
,\quad S_4=\left( \int \!\!d^3x~\alpha _{abcd}\varphi _a\varphi _b\varphi
_c\varphi _d\right) ,  \label{phionly34}
\end{equation}
\begin{equation}
S_5=\int \!\!d^3x~\left( \alpha _{abcde}\varphi _a\varphi _b\varphi
_c\varphi _d\varphi _e\right) ,\quad S_6=\int \!\!d^3x~\left( \alpha
_{abcdef}\varphi _a\varphi _b\varphi _c\varphi _d\varphi _e\varphi _f\right)
,  \label{phionly56}
\end{equation}
where $\alpha ^{\prime }s$ are parameters. The most general invariant local
integrable terms that can be constructed with $B_{\mu \nu }^a$ and $\varphi
^a$ mixed are
\begin{equation}
S_7=\int \!\!d^3x~\left( m_{ab}\epsilon ^{\mu \nu \alpha }H_{a\mu \nu \alpha
}\varphi _b\right) ,\quad S_8=\int \!\!d^3x~\left( m_{abc}\epsilon ^{\mu \nu
\alpha }H_{a\mu \nu \alpha }\varphi _b\varphi _c\right)  \label{mixed12}
\end{equation}
\begin{equation}
S_9=\int \!\!d^3x~\left( m_{abcd}\epsilon ^{\mu \nu \alpha }H_{a\mu \nu
\alpha }\varphi _b\varphi _c\varphi _d\right) ,  \label{mixed3}
\end{equation}
with $m_{ab}$ having dimension of mass, $m_{abc}$ of dimension $1/2$ and $%
m_{abcd}$ as a dimensionless parameter.

Observing the table (\ref{gh1-number-dim}), it is easy to see that
it is impossible to construct invariant local integrated
polynomials with dimension bounded by three with the antifields.
This means that the algebra of the gauge symmetry is undeformed,
i.e., we do not have a nonabelian generalization of the action
(\ref{sa}), the only possibility being with an introduction of
extra fields or non-renormalizable couplings.

Let us now introduce a set of abelian vectorial gauge field in
order to implement the possible nonabelian generalization of
(\ref{sa}). We take the mass dimension of all vector fields equal
to one. Therefore, those fields assume a non-dynamical character.
The BRST transformations are
\begin{equation}
sA_\mu ^a=\partial _\mu c^a,\quad \quad sc^a=0,  \label{nfield}
\end{equation}
where $c^a$ are the ghost for the abelian transformation of $A_\mu
^a$. We must add to the action (\ref{s0}) the corresponding
antifield action
\begin{equation}
S_{ant}^{\prime \prime }=\int \!\!d^3x~A_{a\mu }^{*}\partial ^\mu c_a.
\label{ant-new}
\end{equation}
The new antifields have the following BRST transformations
\begin{equation}
sA_\mu ^{*a}=0,\quad \quad sc^{*a}=\partial _\mu A^{*a\mu }~.
\label{brst-anti}
\end{equation}
We show in the table (\ref{gh2-number-dim}) below the ghost number
and dimension for the new fields (antifields).
\begin{table}[tbh]
\centering
\begin{tabular}{|c|c|c|c|c|}
\hline
& $A_{\mu}^{a}$ & $c^{a}$ & $A^{\ast a}_{\mu}$ & $c^{\ast a}$ \\ \hline
$N_{g}$ & 0 & 1 & -1 & -2 \\ \hline
$dim$ & 1 & 0 & 2 & 3 \\ \hline
\end{tabular}
\caption[t2]{Ghost numbers and dimensions.}
\label{gh2-number-dim}
\end{table}

The all possible invariant integrated local polynomials that can
be constructed with all fields and antifields are
\begin{equation}
S_{10}=g\int \!\!d^3x~f_{abc}\left( \varphi _a^{*}\varphi
_bc_c-\partial ^\mu \varphi _a\varphi _bA_{c\mu }\right),
\end{equation}
\begin{equation}
S_{11}=\mu _{ab}\int \!\!d^3x~\left( A_{a\mu }^{*}\eta _b^\mu
-c_a^{*}\rho _b\right) ,
\end{equation}
\begin{equation}
S_{12}=h\int \!\!d^3x~k_{abc}\left( A_a^{*}A_b^\mu c_c-\frac
12c_a^{*}c_bc_c\right) ,
\end{equation}
where $g,h$ are dimensionless parameter, $\mu $ is a matrix with
dimension 3/2, and $f_{abc}(k_{abc})$ are dimensionless parameters
antisymmetric in its first(last) two indices. Now we perform the
calculation of the antibrackets $(S_i,S_j)$, with $i,j=1,2,\cdots
12$, in order to fit the second order consistency condition. As we
have already seen above, these
antibrackets must be a BRST exact. The antibrackets $(S_m,S_n)$, for $%
n,m=1,2,\dots ,9$ are identically zero, due to absence of antifields in $%
S_n,n=1,2,\dots ,9$. The antibracket $(S_{10},S_{10})$ is
\begin{eqnarray}
(S_{10},S_{10}) &=&g^2\int \!\!d^3x~f_{abc}f_{ab^{\prime }c^{\prime }}\left(
\varphi _{b^{\prime }}^{*}\varphi _bc_cc_{c^{\prime }}+\varphi _{[b}\partial
_\mu \varphi _{b^{\prime }]}A_{c^{\prime }}^\mu c_c\right)  \nonumber \\
&&-\frac{g^2}2s\left( \int \!\!d^3x~f_{abc}f_{ab^{\prime }c^{\prime
}}\varphi _b\varphi _{b^{\prime }}A_{c\mu }A_{c^{\prime }}^\mu \right) ,
\label{s10s10}
\end{eqnarray}
where, $\varphi _{[b}\partial _\mu \varphi _{b^{\prime }]}=\varphi
_b\partial _\mu \varphi _{b^{\prime }}-\varphi _{b^{\prime
}}\partial _\mu \varphi _b$. The first term in (\ref{s10s10}), is
not a BRST trivial and it could jeopardize the nonabelian
implementation. In order to circumvent this, we must have the
identification $hk_{abc}=gf_{abc}$, and $f_{abc}$ being the
structure constant of a Lie group. Therefore the $S_{10}$ and
$S_{12}$ are replaced by the sum
\begin{equation}
S_{10}^{\prime }=g\int \!\!d^3x~f^{abc}\left( \varphi ^{*a}\varphi
^bc^c-\partial ^\mu \varphi ^a\varphi ^bA_\mu ^c+A^{*a}A^{b\mu }c^c-\frac
12c^{*a}c^bc^c\right) .  \label{S10l}
\end{equation}
It is easy to see that now $(S_{10}^{\prime },S_{10}^{\prime })$ is BRST
trivial
\begin{equation}
(S_{10}^{\prime },S_{10}^{\prime })=-\frac{g^2}2s\left( \int
\!\!d^3x~f_{abc}f_{ab^{\prime }c^{\prime }}\varphi _b\varphi _{b^{\prime
}}A_{c\mu }A_{c^{\prime }}^\mu \right) .  \label{sl10sl10}
\end{equation}
The antibrackets $(S_{10}^{\prime },S_n)$, with $n=1,2,\dots ,6$,
give us constraint for the parameters $\alpha $: $\alpha _a=\alpha
_{abc}=\alpha _{abcde}=0$, $\alpha _{ab}=a_1\delta _{ab}$, $\alpha
_{abcd}=a_2\delta _{ab}\delta _{cd}$, $\alpha _{abcdef}=a_3\delta
_{ab}\delta _{cd}\delta _{ef} $, i.e., only the terms $\varphi
^2=\varphi _a\varphi _a$, $(\varphi ^2)^2$ and $(\varphi ^2)^3$
are permitted. The last antibrackets reads
\[
(S_{11},S_{11})=0,
\]
\begin{eqnarray}
(S_{10}^{\prime },S_{11})=g\int \!\!d^3x~\!\!\!\! &&f_{abc}\mu _{ab^{\prime
}}\left( \rho _{b^{\prime }}\varphi _b^{*}\varphi _c+\rho _{b^{\prime
}}A_{b\mu }^{*}A_c^\mu -\rho _{b^{\prime }}c_b^{*}c_c\right.  \nonumber \\
&&\left. -\eta _{b^{\prime }}^\mu \partial _\mu \varphi _b\varphi _c-\eta
_{b^{\prime }}^\mu A_{b\mu }^{*}c_c\right) ,  \label{s11}
\end{eqnarray}
\[
(S_{10}^{\prime },S_7)=g\int \!\!d^3x~f_{abc}m_{b^{\prime }a}\varepsilon
_{\mu \nu \alpha }H_{b^{\prime }}^{\mu \nu \alpha }\varphi _bc_c,
\]
\[
(S_{10}^{\prime },S_8)=g\int \!\!d^3x~f_{abc}(m_{b^{\prime }c^{\prime
}a}+m_{b^{\prime }ac^{\prime }})\varepsilon _{\mu \nu \alpha }H_{b^{\prime
}}^{\mu \nu \alpha }\varphi _b\varphi _{c^{\prime }}c_c,
\]
\[
(S_{10}^{\prime },S_9)=g\int \!\!d^3x~f_{abc}(m_{b^{\prime }c^{\prime
}d^{\prime }a}+m_{b^{\prime }c^{\prime }ad^{\prime }}+m_{b^{\prime
}ac^{\prime }d^{\prime }})\varepsilon _{\mu \nu \alpha }H_{b^{\prime }}^{\mu
\nu \alpha }\varphi _b\varphi _{c^{\prime }}\varphi _{d^{\prime }}c_c.
\]
The last four antibrackets are not BRST trivial, representing thus
an obstruction to the deformation of the master equation. The only
way to remedy this is setting $g=0$, or setting
$S_7=S_8=S_9=S_{11}=0$. In the case $g=0$ we have lost the
deformation of the abelian algebra, i.e, we have a set of abelian
fields not representing a nonabelian generalization of (\ref
{sa}). In the case in which $S_7=0$, we have lost the mass
generation of the model. We have thus proved that there are no
nonabelian generalization of the action (\ref{sa}), even with an
addition of an auxiliary vector gauge field.

It is interesting to remark that the introduction of an one form
gauge connection $A$ is required to go further in the nonabelian
generalization of our model (\ref{sa}), although our original
abelian action (\ref{sa}) does not contain this field.

We can quote some works that built some kind of nonabelianization
of the model under consideration. However, the work of Oda and
Yahikozawa \cite{oda} as well as the work of Smailagic and
Spallucci \cite{smailagic}, have considered the connection one
form $A$ as a flat background field (so imposing an extra
constraint in the model). On the other hand, Del Cima {\it et al.}
\cite{delcima}, have studied the finiteness of a three-dimensional
extension of the BF model, called BFK model, which does not have
kinetic term for the antisymmetric gauge field (this term violate
the nonabelian gauge invariance which is straightforwardly
implemented for vectorial fields).

Motivated by possible nonabelian topological mass generation for a
Kalb-Ramond field in three dimensions, we have considered
deformations of a model involving a topological coupling between a
second rank antisymmetric tensor field and a scalar field, using
the method of consistent deformations. But an obstruction arise,
leading us to a no-go theorem, namely, if we require
power-counting renormalizable couplings and the same field
content, the nonabelian extension for the model is forbidden.

\vspace{0.3in} \centerline{\bf ACKNOWLEDGMENTS}

We would like to thank Dr. O. S. Ventura for helpful discussions.
This work was supported in part by Funda\c{c}\~{a}o Cearense de
Amparo \`{a} Pesquisa-FUNCAP.

\end{document}